\documentclass[article,12pt]{article} 

\usepackage{amscd,amsmath,amssymb,amsfonts,latexsym,mathrsfs,amsthm,mathtools}
\usepackage{graphicx} 
\usepackage{setspace} 
\usepackage{graphics,epsfig}
\graphicspath{{figures/}} 
\usepackage{sectsty}
\usepackage[numbers]{natbib}
\usepackage[english]{babel}
\usepackage{amsmath}
\usepackage{amsfonts}
\usepackage{amssymb,amsthm}
\usepackage{bm}
\usepackage{mathrsfs}
\usepackage{subcaption}
\usepackage[usenames]{color}
\usepackage{rotating}
\usepackage{url}
\usepackage{bbm}
\usepackage{float}

\UseRawInputEncoding

\newtheorem{rem}{{\bf Remark}}

\usepackage{flushend}
\usepackage{verbatim}
\usepackage{tabularx}
\usepackage{multirow} 
\usepackage{arydshln}
\usepackage{hyperref}
\usepackage[toc,page]{appendix}
\usepackage{bigints}

\newcommand{\E}{\mathbb{E}}

\newcommand{\x}{{\bf x}}
\newcommand{\z}{{\bf z}}

\newcommand{\y}{{\bf y}}

\newcommand{\post}{\bar{\pi}}

\title{Target-aware Bayesian inference via generalized thermodynamic integration}

\author{F. Llorente$^\dagger$, L. Martino$^{\star}$, D. Delgado$^\dagger$ \\
{\small$^\dagger$  Universidad Carlos III de Madrid,  Legan\'es (Spain).}\\
{\small$^\star$  Universit{\'a} degli Studi di Catania,  Italy.} \\
}

\date{}
 
\begin{document}

\maketitle

\begin{abstract}

In Bayesian inference, we are usually interested in the numerical approximation of  integrals that are posterior expectations or marginal likelihoods (a.k.a., Bayesian evidence).
In this paper, we focus on the computation of the posterior expectation of a function $f(\x)$. 
We consider a  \emph{target-aware} scenario where $f(\x)$ is known in advance and can be exploited in order to improve the estimation of the posterior expectation.
In this scenario, this task can be reduced to perform several independent marginal likelihood estimation tasks.
 The idea of using a path of tempered posterior distributions has been widely applied in the literature for the computation of marginal likelihoods.   Thermodynamic integration, path sampling and annealing importance sampling are well-known examples of algorithms belonging to this family of methods.  In this work, we introduce a generalized thermodynamic integration (GTI) scheme which is able to perform a target-aware Bayesian inference, i.e.,  GTI can approximate the posterior expectation of a given function. Several scenarios of application of GTI are discussed and different numerical simulations are provided.
\end{abstract}


%
%
%
%
%
%

\section{Introduction}

Bayesian methods have become very popular in many domains of science and engineering over the last years, as they allow for obtaining estimates of parameters of interest as well as comparing competing models in a principled way \cite{Robert04,LuengoMartino2020}.  The Bayesian quantities can generally be expressed as integrals involving the posterior density.  They can be divided in two main categories: posterior expectations and marginal likelihoods (useful for model selection purposes). 
\newline
Generally, computational methods are required for the approximation of these integrals, e.g., Monte Carlo algorithms such as Markov chain Monte Carlo (MCMC) and importance sampling (IS) \cite{Robert04,LuengoMartino2020,rainforth2020target}.  
{
	Typically, practitioners apply an MCMC or IS algorithm to approximate the posterior $\post(\x)$ density by a set of samples, which is used in turn to estimate posterior expectations $\E_{\post}[f(\x)]$ of some function $f(\x)$.  
	Although it this is a sensible strategy when  $f(\x)$ is not known in advance and/or we are interested in computing several posterior expectations with respect to different functions, this strategy is suboptimal when the \emph{target} function $f(\x)$ is known in advance since it is completely agnostic to $f(\x)$ . 
Incorporating knowledge of $f(\x)$ for the estimation of $\E_{\post}[f(\x)]$ is known as \emph{target-aware Bayesian inference} or TABI \cite{rainforth2020target}.
TABI proposes to break down the estimation of the posterior expectation into several independent estimation tasks. Specifically, in TABI we require to estimate three marginal likelihoods (or normalizing constants) independently, and then recombine the estimates in order to form the approximation of the posterior expectation. 
The target function $f(\x)$ features in two out of the three marginal likelihoods that have to be estimated. Hence, the TABI framework provides means of improving the estimation of a posterior expectation by making use explicitly of $f(\x)$ and leveraging the use of algorithms for marginal likelihood computation.
}
\newline
The computation of the marginal likelihoods is particularly complicated, specially with MCMC outputs \cite{newton1994approximate,llorente2020marginal,llorente2021computation}.
{
IS techniques are the most popular for this task. The basic IS algorithm provides with a straightforward estimation of the marginal likelihood. However, designing a good proposal pdf that approximates the target density is not easy \cite{llorente2020marginal}.
}
For this reason,  sophisticated and powerful schemes  have been specifically designed  \cite{llorente2020marginal,friel2012estimating}. The most powerful techniques involve  the idea of  the so-called {\it tempering} of the posterior \cite{Neal01,lartillot2006computing,friel2008marginal}.
The tempering effect is commonly employed in order to foster the exploration and improve the efficiency of MCMC chains \cite{neal1996sampling,martino2021automatic}. 
State-of-the-art methods for computing marginal likelihoods consider tempered transitions (i.e. sequence of tempered distributions), such as annealed IS (An-IS) \cite{Neal01}, sequential Monte Carlo (SMC) \cite{Moral06}, thermodynamic integration (TI), a.k.a., path sampling (PS) or  ``power posteriors'' (PP) in the statistics literature \cite{lartillot2006computing,friel2008marginal,gelman1998simulating}, and stepping stones (SS) sampling \cite{xie2010improving}. 
An-IS is a special case of SMC framework, PP is a special case of TI/PS, and SS sampling present similar features to An-IS and PP. For more details, see \cite{llorente2020marginal}. It is worth to mention that TI has been introduced in the physics literature for computing free-energy differences \cite{frenkel1986free,gelman1998simulating}.  

\noindent In this work, we extend the TI method, introducing the generalized thermodynamic integration (GTI) technique, for computing posterior expectations of a function $f(\x)$. 
{In this sense, GTI is a target-aware algorithm that incorporates information $f(\x)$ within the marginal likelihood estimation technique TI.}
The extension of TI for the computation of $\E_{\post}\left[f(\x)\right]$ is not straightforward, since it requires to build a continuous path between densities with possibly different support. In the case of a geometric path (which is the default choice in practice \cite{friel2008marginal,lartillot2006computing}), the generalization of TI needs a careful look at the support of the negative and positive parts of $f(\x)$.  
We discuss the application of GTI for the computation of posterior expectations of generic real-valued function $f(\x)$, and also describe the case of vector-valued function ${\bf f}(\x)$. 
The benefits of GTI are clearly shown by illustrative numerical simulations.
\newline
The structure of the paper is the following. In Section \ref{sec_background}, we introduce the Bayesian inference setting and describe the thermodynamic method for the computation of the marginal likelihood. In Section \ref{sec_GTI}, we introduce the GTI procedure. More specifically, we discuss first the case when $f(\x)$ is strictly {positive or negative} in Section \ref{sec_f_pos}, and then consider the general case of a real-valued $f(\x)$ in Section \ref{sec_f_gen}. In Section \ref{sec_compDet}, we discuss some computational details of the approach, and the application of GTI for vector-valued functions ${\bf f}(\x)$.
We show the benefits of GTI in two numerical experiments in Section \ref{sec_exper}. Finally, Section \ref{sec_conc} contains the conclusions.

\section{Background}\label{sec_background}

\subsection{Bayesian inference}\label{sec_BI}
In many real-world applications, the goal is to infer a {parameter} of interest given a set of data \cite{Robert04}.
Let us denote the parameter of interest by $\x\in \mathcal{X}\subseteq \mathbb{R}^{D}$, and let ${\bf y}\in \mathbb{R}^{d_y}$ be the observed data. In a Bayesian analysis, all the statistical information is contained in the posterior distribution, which is given by
\begin{equation}
\post(\x)= p(\x|\y)= \frac{\ell(\y|\x) g(\x)}{Z(\y)},
\label{eq:posterior}
\end{equation}
where $\ell(\y|\x)$ is the likelihood function, $g(\x)$ is the prior pdf, and $Z(\y)$ is the  Bayesian model evidence (a.k.a. marginal likelihood). 
%
Generally, $Z(\y)$ is unknown, so we are able to evaluate the unnormalized target function,
$\pi(\x)=\ell(\y|\x) g(\x)$. The analytical computation of the posterior density $\post(\x) \propto \pi(\x)$ is often unfeasible, hence numerical approximations are needed. The interest lies in in the approximation of integrals of the form 
\begin{align}\label{eq:IntegralOfInter}
I = \E_{\post}\left[f(\x)\right]= \int_\mathcal{X}f(\x)\post(\x)d\x =\frac{1}{Z} \int_\mathcal{X}f(\x)\pi(\x)d\x,
\end{align}
where $f(\x)$ is some integrable function, and 
\begin{align}\label{eq:MargLike}
Z = \int_\mathcal{X}\pi(\x)d\x.
\end{align}
The quantity $Z$ is called {\it marginal likelihood} (a.k.a., {\it Bayesian evidence}) and is useful for model selection purpose \cite{llorente2020marginal}.
Generally, $I$ are $Z$ analytically intractable and we need to resort to numerical algorithms such as Markov chain Monte Carlo (MCMC) and importance sampling (IS) algorithms.
{
In this work, we consider $f(\x)$ is known in advance, and we aim at exploiting it in order to apply thermodynamic integration for computing the posterior expectation $I$, namely, perform \emph{target-aware} Bayesian inference (TABI). 
}

{
	
	\subsection{Computation of marginal likelihoods for parameter estimation: The TABI framework}
	
%
%
	
	The focus of this work is on parameter estimation, namely, we are interested in the computation of the posterior expectation in Eq. \eqref{eq:IntegralOfInter} of some function $f(\x)$.
	Recently, the authors in \cite{rainforth2020target} propose a framework called \emph{target-aware Bayesian inference} (TABI) that aims at improving the Monte Carlo estimation of $I$ when the \emph{target} $f(\x)$ is known in advance. The TABI framework is based on decomposing $I$ into several terms and estimate them separately, leveraging the information in $f(\x)$. Let $f_+(\x) = \max\{0,f(\x)\}$ and $f_-(\x) = \max\{0,-f(\x)\}$, so $f(\x) = f_+(\x) - f_-(\x)$. 	
	Hence, TABI rewrites the posterior expectation $I$ as
	\begin{align}\label{eq:TABIidentity}
		I = \frac{c_+ - c_-}{Z},
	\end{align}
	where $c_+ = \int f_+(\x)\pi(\x)d\x$ and $c_- = \int f_-(\x)\pi(\x)d\x$. Note that $c_+$, $c_-$ and $Z$ are integrals of non-negative functions, namely, they are marginal likelihoods (or normalizing constants). 
	The three unnormalized densities of interest hence are $\pi(\x)$, $f_+(\x)\pi(\x)$ and $f_-(\x)\pi(\x)$. 
	Note that two out of the three (unnormalized) densities incorporate information about $f(\x)$.
	The general TABI estimator is then
	\begin{align}\label{eq:TABIestimator}
		\widehat{I}_{\text{TABI}} = \frac{\widehat{c}_+ - \widehat{c}_-}{\widehat{Z}},
	\end{align}
	where  $\widehat{c}_+$, $\widehat{c}_-$ and $\widehat{Z}$ are estimates obtained independently. These estimates can be obtained by any marginal likelihood estimation method. The original TABI framework is motivated in the IS context.
	This is due to the fact that marginal likelihoods (i.e., integrals of non-negative functions) can be estimated arbitrarily well with IS \cite{llorente2020marginal,llorente2021computation,rainforth2020target}. Namely, using the optimal proposals the estimates $\widehat{c}_+$, $\widehat{c}_-$ and $\widehat{Z}$ coincide with the exact values regardless of the sample size. Note that the direct estimation of $I$ via MCMC or IS cannot produce zero-variance estimators for a finite sample size \cite{Robert04}. 
	\newline
	\newline
	The TABI framework improves the estimation of $I$ by converting the initial task in that of computing three marginal likelihoods, $c_+$, $c_-$ and $Z$. 
	In \cite{rainforth2020target}, the authors test the application of two popular marginal likelihood estimators within TABI, namely, annealed IS (AnIS) \cite{neal1996sampling} and nested sampling (NS) \cite{skilling2006nested}, resulting in the target-aware algorithms called \emph{target-aware} AnIS (TAAnIS) and \emph{target-aware} NS (TANS). 
	The use of AnIS for independently computing $c_+$, $c_-$ and $Z$ represents an improvement over IS. Although the IS estimation of $c_+$, $c_-$ and $Z$ can have virtually zero-variance, this is only true when we employ the optimal proposals. In general, the performance of IS depends on how `close' is the proposal pdf to the target density whose normalizing constant we aim to estimate. It can be shown that the variance of IS scales with the Pearson divergence between target and proposal \cite{llorente2020marginal}. When this distance is large, then it is more efficient to sample from another proposal that is `in between', i.e., an `intermediate' density . This is the motivation behind many state-of-the-art marginal likelihood estimation methods that employ a sequence of densities bridging an easy-to-work-with proposal and the target density \cite{llorente2020marginal}. 
	In this work, we introduce thermodynamic integration (TI) for performing target-aware inference, hence enabling the computation of the posterior expectation of a function $f(\x)$. 
	TI is a powerful marginal likelihood estimation technique that also leverages the use of a sequence of distributions, but has several advantages over other methods based on tempered transitions, such as improved stability thanks to working in logarithm scale and applying deterministic quadrature \cite{friel2008marginal,friel2012estimating}. 
	TI for computing marginal likelihoods is reviewed in the next section. Then, in Sect. \ref{sec_GTI} we introduce generalized TI (GTI) for the computation of posterior expectations, that is based on rewriting $I$ as the difference of two ratios of normalizing constants. 

}

\subsection{Thermodynamic integration for estimating $Z$}\label{sec_TI}
Thermodynamic integration (TI) is a powerful technique that has been proposed in literature for computing ratios of constants \cite{frenkel1986free,gelman1998simulating,lartillot2006computing}. Here, for simplicity, we focus on the approximation of just one constant, the marginal likelihood $Z$. More precisely, TI produces an estimation of $\log Z$.  Let us consider a family of (generally unnormalized) densities 
\begin{align}
\pi(\x|\beta), \quad \beta\in[0,1],
\end{align}
such that $\pi(\x|0)=g(\x)$ is the prior and $\pi(\x|1)=\pi(\x)$ is the unnormalized posterior distribution. An example is  the so-called {\it geometric path} $\pi(\x|\beta) = g(\x)^{1-\beta}\pi(\x)^\beta$, with $\beta\in[0,1]$ \cite{neal1993probabilistic}.
The corresponding normalized densities in the family are denoted as
\begin{align}
\bar{\pi}(\x|\beta) = \frac{\pi(\x|\beta)}{c(\beta)},\quad c(\beta) = \int_{\mathcal{X}} \pi(\x|\beta)d\x.
\end{align}
Then, the main TI identity is \cite{llorente2020marginal}
\begin{align}
	\log Z &=\int_0^1 \left[ \int_\mathcal{X} \frac{\partial \log \pi(\x|\beta)}{\partial \beta}\post(\x|\beta) d\x\right]d\beta  \nonumber \\
	&= \int_0^1 \E_{\post(\x|\beta)}\left[\frac{\partial \log \pi(\x|\beta)}{\partial \beta}\right]d\beta, \label{eq_thermo_gen}
\end{align}
where the expectation is with respect to (w.r.t.)  $\post(\x|\beta) = \frac{\pi(\x|\beta)}{c(\beta)}$. 
\newline
{\bf  TI estimator.} Using an ordered sequence of discrete values  $\{\beta_i\}_{i=1}^N$ (e.g. $\beta_i$'s uniformly in $[0,1]$), one can approximate the integral in Eq. \eqref{eq_thermo_gen} via quadrature w.r.t. $\beta$, and then approximate the inner expectation with a Monte Carlo estimator using samples from $\post(\x|\beta_i)$ for $i=1,\dots,N$.  Namely, defining $U(\x) = \frac{\partial \log \pi(\x|\beta)}{\partial \beta}$ and $E(\beta) =\E_{\post(\x|\beta)}\left[U(\x)\right]$,
the resulting estimator of Eq. \eqref{eq_thermo_gen} is given by
\begin{align}\label{eq_thermo_est}
	\log Z \approx \sum_{i=1}^N(\beta_{i+1} - \beta_{i})\widehat{E}_{i},
\end{align}
where 
\begin{align}\label{eq_MonteCarloEstims}
	\widehat{E}_i = \frac{1}{N}\sum_{j=1}^N U(\x_{i,j}),\quad \x_{i,j} \sim p(\x|\beta_i).
\end{align}
Note that we used the simplest quadrature rule in Eq. \eqref{eq_thermo_est}, but others can be used such as Trapezoidal, Simpson's, etc \cite{friel2008marginal,lartillot2006computing}.
\newline
{\bf The power posteriors (PP) method.}  Let us consider the specific case of a geometric path between prior $g(\x)$ and unnomalized posterior $\pi(\x)$,
\begin{align}
\pi(\x|\beta) = g(\x)^{1-\beta}\pi(\x)^\beta &=g(\x) \left[\frac{\pi(\x)}{g(\x)}\right]^\beta, \\
&=g(\x) \ell(\y |\x)^\beta,  \quad \beta\in[0,1], \label{aquiBuenPI}	
\end{align}
where we have used $\pi(\x)=\ell(\y |\x) g(\x)$. 
Note that, in this scenario,\footnote{From Eq. \eqref{aquiBuenPI}, we can write $\log \pi(\x|\beta)= \log g(\x)+\beta \log \ell(\y |\x)$. Hence, $\frac{\partial \log \pi(\x|\beta)}{\partial \beta}=\log \ell(\y |\x)$.}
\begin{align}
\frac{\partial \log \pi(\x|\beta)}{\partial \beta}
=\log \ell(\y |\x).
\end{align}
Hence, the identity in Eq. \eqref{eq_thermo_gen} can be also written as
\begin{align}\label{eq_basic_PP}
	\log Z = \int_0^1 \int_\mathcal{X} \log \ell(\y|\x) \post(\x|\beta)d\x d\beta = \int_0^1\E_{\post(\x|\beta)}[\log \ell(\y|\x)]d\beta,
\end{align}
The {\it power posteriors (PP) method} is a special case of TI which considers {\bf (a)} the geometric path and  {\bf (b)} trapezoidal quadrature rule for integrating w.r.t. the variable $\beta$ \cite{friel2008marginal}. Namely, letting $\beta_1=0 < \dots < \beta_N = 1$ denote a fixed temperature schedule, an approximation of Eq. \eqref{eq_basic_PP} can be obtained via the trapezoidal rule
\begin{align}\label{eq_basic_PP_estimator}
	\log Z \approx \sum_{i=1}^{N-1} (\beta_{i+1}-\beta_i)\frac{\E_{\post(\x|\beta_{i+1})}[\log \ell(\y|\x)]+\E_{\post(\x|\beta_{i})}[\log \ell(\y|\x)]}{2},
\end{align}
where the the expectations are generally substituted with MCMC estimates as in Eq. \eqref{eq_MonteCarloEstims}. 
TI and PP are popular methods for computing marginal likelihoods (even in high-dimensional spaces) due to their reliability. Theoretical properties are studied in \cite{gelman1998simulating,calderhead2009estimating}, and empirical validation is provided in several works, e.g.,  \cite{friel2008marginal,lartillot2006computing}. 
Different extensions and improvements on the method have also been proposed \cite{oates2016controlled,friel2014improving,calderhead2009estimating}.                 
{\rem\label{Rem1} Note that, in order to ensure that the integrand in Eq. \eqref{eq_basic_PP} is finite, so that the estimator in Eq. \eqref{eq_basic_PP_estimator} can be applied, we need that (a) $\ell(\y|\x)$ is strictly positive everywhere, or (b) $\ell(\y|\x)=0$ only whenever $g(\x)=0$ (i.e., they have the same support). 
}
\newline
\newline
{\bf Goal}. We have seen that the TI method has been proposed for computing $\log Z$ (or log-ratios of constants). Our goal is to extend the TI scheme in order to perform target-aware Bayesian inference.
Namely, we generalize the idea of these methods (thermodynamic integration, power posteriors, etc.) to the computation of   posterior expectations for a given $f(\x)$. 


\section{Generalized TI (GTI) for Bayesian inference}\label{sec_GTI}

In this section, we extend the TI method for computing the posterior expectation of a given $f(\x)$. 
As in TABI, the basic idea, as we show below, is the formulation $I$ in terms of ratios of normalizing constants.  
First, we consider the case $f(\x)> 0$ for all $\x$ and then the case of a generic real-valued $f(\x)$.

\subsection{General approach}\label{sec_tres_dens}


In order to apply TI, we need to formulate the posterior expectation $I$ as a ratio of two constants. Since $f(\x) $ can be positive or negative, 
let us consider the positive and negative parts, $f_+(\x) = \max(0,f(\x))$ and $f_-(\x) = \min(0,-f(\x))$, such that $f(\x) = f_+(\x) - f_-(\x)$, where $f_+(\x)$ and $f_-(\x)$ are non-negative functions.
Similarly to Eq. \eqref{eq:TABIidentity}, we rewrite the integral $I$ in terms of ratios of constants,
\begin{align}\label{eq_3_const}
I = \dfrac{ \int_{\mathcal{X}}  f_+(\x)\pi(\x)d\x}{\int_{\mathcal{X}}  \pi(\x)d\x} - \dfrac{\int_{\mathcal{X}}  f_-(\x)\pi(\x)d\x}{\int_{\mathcal{X}}  \pi(\x)d\x}
=\frac{c_+}{Z} - \frac{c_-}{Z},
\end{align}
where $c_+= \int_{\mathcal{X}} \varphi_+(\x)d\x$ are $c_-= \int_{\mathcal{X}} \varphi_-(\x)d\x$ are respectively the normalizing constants of $\varphi_+(\x) = f_+(\x)\pi(\x)$, and 
$\varphi_-(\x) = f_-(\x)\pi(\x)$.
\newline
{\bf Proposed scheme.} Denoting  $\eta_+ = \log\frac{c_+}{Z}$ and $\eta_- = \log\frac{c_-}{Z}$ in the case of a generic $f(\x)$, we propose to obtain estimates of these quantities using thermodynamic integration.
Then, we can obtain the final estimator as
\begin{align}\label{eq_luca}
\widehat{I} = \exp\left(\widehat{\eta}_+\right) - \exp\left(\widehat{\eta}_-\right).
\end{align}
In the next section, we give details on how to compute $\widehat{\eta}_+$, $\widehat{\eta}_-$ by using a generalized TI method. 

{\rem Note that in Eq. \eqref{eq_3_const} we express $I$  as the difference of two ratios, and we propose GTI to estimate them directly as per Eq. \eqref{eq_luca} Hence, differently from Eq. \eqref{eq:TABIestimator}, we do not aim at estimating each constant separately. This amounts to bridging the posterior with the function-scaled posterior, as we show below.
}

\subsection{GTI for strictly positive or strictly negative  $f(\x)$}\label{sec_f_pos}

Let us consider the scenario where $f(\x)>0$ for all $\x\in\mathcal{X}$. 
In this scenario, we can set
 $$
 f_+(\x)=f(\x)>0,\quad \mbox{and} \quad \widehat{I} = \exp\left(\widehat{\eta}_+\right).
 $$ 
 Note that, with respect to Eq. \eqref{eq_luca}, we only consider the first term.
We link the unnormalized pdfs $\pi(\x)$ and $\varphi_+(\x)=f_+(\x)\pi(\x)$ with a geometric path, by defining
\begin{align}
\bar{\varphi}_+(\x|\beta) \propto \varphi_+(\x|\beta)= f_+(\x)^{\beta}\pi(\x),\quad \beta\in[0,1].
\end{align}
Hence, we have $\bar{\varphi}_+(\x|0)=\post(\x)$ and $\bar{\varphi}_+(\x|1) = \frac{1}{c_+}f_+(\x)\pi(\x)$ where $c_+=\int_{\mathcal{X}} f_+(\x)\pi(\x) d\x$. 
The Eq. \eqref{eq_thermo_gen} is thus
\begin{align}\label{eq_f_strictly}
\eta_+ =  \int_0^1 \E_{\bar{\varphi}_+(\x|\beta)}[\log f_+(\x)]d\beta.
\end{align}
Letting $\beta_1=0 < \dots < \beta_N = 1$ denote a fixed temperature schedule, the estimator (using the Trapezoidal rule) is thus
\begin{align}\label{eq_TIquad}
	\widehat{\eta}_+ = \sum_{i=1}^{N-1} (\beta_{i+1}-\beta_i)\frac{\E_{\bar{\varphi}_+(\x|\beta_{i+1})}[\log f_+(\x)]+\E_{\bar{\varphi}_+(\x|\beta_i)}[\log f_+(\x)]}{2},
\end{align}
where we use MCMC estimates for the terms
\begin{align}
	\E_{\bar{\varphi}_+(\x|\beta_i)}[\log f_+(\x)] = \int_\mathcal{X}\log f_+(\x)\bar{\varphi}_+(\x|\beta_i)d\x \approx \frac{1}{M}\sum_{m=1}^M \log f_+(\x_m),\quad \x_m \sim \bar{\varphi}_+(\x|\beta_i),
\end{align}
for $i=1,\dots,N$.
The case of a strictly negative $f(\x)$, i.e., $f_-(\x)=-f(\x)$, is equivalent.
\newline
\newline
{\bf Function $f(\x)$ with zeros with null measure.} So far, we have considered strictly positive or strictly negative  $f(\x)$. This case could be extended to a positive (or negative)  $f(\x)$ with zeros in a  null measure set.
Indeed, note that the identity in Eq. \eqref{eq_f_strictly} requires that $\E_{\bar{\varphi}(\x|\beta)}[\log f(\x)]<\infty$ for all $\beta\in[0,1]$. If the zeros of $f(\x)$ has null measure and the improper integral converges, the procedure above is also suitable.
Table \ref{table_f_stricly_pos} summarizes the Generalized TI (GTI) steps for $f(\x)$ that are strictly positive.
We discuss other scenarios in the next section.

\begin{table}[!ht]
	\caption{\textbf{GTI for strictly positive $f(\x)$} }
	\vspace{-0.2cm}
	\begin{tabular}{|p{0.95\columnwidth}|}
		\hline
		\vspace{0.1cm}
		{\bf - Initialization:} Choose the set of nodes $\{\beta_i\}_{i=1}^N$ (with $\beta_1=0$ and $\beta_N=1$), and the number of iterations $N$.
		\newline
		{\bf - For $i=1,\ldots,N$:}
		\begin{enumerate}
			\item  {\bf Sampling:} Sample $\{\x_{i,m}\}_{m=1}^M\sim \bar{\varphi}_+(\x|\beta_i) \propto f(\x)^{\beta_i}\pi(\x)$.
			\item {\bf Compute:} 
			\begin{align}
			\widehat{E}_i = \frac{1}{M}\sum_{k=1}^{M}\log f(\x_{i,k}).
			\end{align}
			\item {\bf Update:} If $i=1$, set $\eta^{(i)}=0$. If $i>1$, update recursive estimate
			\begin{align}
				\widehat{\eta}^{(i)} \leftarrow \widehat{\eta}^{(i-1)} + 0.5\cdot(\beta_{i}-\beta_{i-1})\left(\widehat{E}_i+\widehat{E}_{i-1}\right).
			\end{align}
		\end{enumerate}
		\vspace{-0.3cm}
		{\bf - Outputs:} Final estimator $\widehat{\eta}^{(N)}$ which approximates $\log I$. 
		\\ 
		\hline 
	\end{tabular}
	\label{table_f_stricly_pos}
\end{table}

\subsection{GTI for generic $f(\x)$}\label{sec_f_gen}

Using the results from previous section, we apply GTI to a real-valued function $f(\x)$, namely, it can be positive and negative, as well as having zero-valued regions with a non-null measure. Here, we desire to connect the posterior $\pi(\x)$ with  the $f_+(\x)\pi(\x)$ and $f_-(\x)\pi(\x)$ with two continuous paths. However,
a requirement for the validity of the approach is that $\pi(\x)$ is zero whenever $f_+(\x)\pi(\x)$ or $f_-(\x)\pi(\x)$ is zero, which does not generally fulfills as $f(\x)$ can have a smaller support than $\pi(\x)$. 
This fact enforces the computation of correction factors to keep the validity of the approach. More details can be found in App. \ref{App:ContinuousPath}.
Therefore, we need to define the unnormalized restricted posteriors densities
\begin{align}
	\pi_+(\x) = \pi(\x)\mathbbm{1}_{\mathcal{X}_+}(\x),\quad \text{and} \quad \pi_-(\x) = \pi(\x)\mathbbm{1}_{\mathcal{X}_-}(\x),
\end{align}
where $\mathbbm{1}_{\mathcal{X}_+}(\x)$ is the indicator function over the set 
$\mathcal{X}_+ = \{\x\in\mathcal{X}: f_+(\x)>0\}$ and $\mathbbm{1}_{\mathcal{X}_-}(\x)$   is the indicator function over the set $\mathcal{X}_- = \{\x\in\mathcal{X}: f_-(\x)>0\}$.
The idea is to connect with a path $\pi_+(\x)$ and $f_+(\x)\pi(\x)$, and $\pi_-(\x)$ with $f_-(\x)\pi(\x)$, by the densities
\begin{align*}
\bar{\varphi}_+(\x|\beta) \propto f_+(\x)^{\beta}\pi_+(\x), \qquad \bar{\varphi}_-(\x|\beta) \propto f_-(\x)^{\beta}\pi_-(\x), \quad \beta\in[0,1].
\end{align*}
Note that it is equivalent to write $f_{\pm}(\x)\pi_{\pm}(\x) = f_{\pm}(\x)\pi(\x)$,  since $\pi_{\pm}(\x) = \pi(\x)$ whenever $f_{\pm}(\x)>0$, and they only differ when $f_{\pm}(\x) = 0$, in which case we also have $f_{\pm}(\x)\pi_{\pm}(\x) = f_{\pm}(\x)\pi(\x) = 0$.
 Defining  also
\begin{align}
	Z_+=\int_{\mathcal{X}}\pi_+(\x) d\x, \quad Z_-=\int_{\mathcal{X}}\pi_-(\x) d\x, 
\end{align}
and recalling
\begin{align}
	c_+=\int_{\mathcal{X}}  f_+(\x)\pi(\x)d\x, \quad c_-=\int_{\mathcal{X}}  f_-(\x)\pi(\x)d\x, 
\end{align}
the idea is to  apply separately TI for approximating $\eta^\text{res}_+=\log\frac{c_+}{Z_+}$ and $\eta^\text{res}_-=\log\frac{c_-}{Z_-}$, where we denote with {\it res} to account that we consider the restricted components $Z_+$ and $Z_-$. Hence, two correction factors $R_+$ and $R_{-}$ are also required, in order to obtain $R_+\exp\left({\eta^\text{res}_+}\right)=\frac{c_+}{Z}$ and $R_{-}\exp\left({\eta^\text{res}_-}\right)=\frac{c_-}{Z}$.
Below, we also show how to estimate the correction factors at a final stage and combine them to the estimations of $\eta^\text{res}_+$ and $\eta^\text{res}_-$.
We can approximate the quantities 
\begin{align*}
	\eta^\text{res}_+ = \log \frac{c_+}{Z_+} = \int_0^1 \E_{\bar{\varphi}_+(\x|\beta)}[\log f_+(\x)]d\beta, \\
	\eta^\text{res}_- = \log \frac{c_-}{Z_-} = \int_0^1 \E_{\bar{\varphi}_-(\x|\beta)}[\log f_+(\x)]d\beta,
\end{align*}
 using the estimators
\begin{align}\label{eq_res_pos}
	\widehat{\eta}^\text{res}_+ = \sum_{i=1}^{N-1}(\beta_{i+1} - \beta_{i})\frac{\widehat{E}^+_{i+1} + \widehat{E}^+_i}{2}, \\
	\widehat{\eta}^\text{res}_- = \sum_{i=1}^{N-1}(\beta_{i+1} - \beta_{i})\frac{\widehat{E}^-_{i+1} + \widehat{E}^-_i}{2}, \label{eq_res_neg}
\end{align}
where 
\begin{align}
\widehat{E}^+_i = \frac{1}{M}\sum_{m=1}^{M} \log f_+(\x_{i,m}),\quad \x_{i,m} \sim \bar{\varphi}_+(\x|\beta_i), \\
\widehat{E}^-_i = \frac{1}{M}\sum_{m=1}^{M} \log f_-({\bf v}_{i,m}),\quad {\bf v}_{i,m} \sim \bar{\varphi}_-(\x|\beta_i).
\end{align}
When comparing the estimators in Eqs. \eqref{eq_res_pos}-\eqref{eq_res_neg} with respect to the GTI estimator in Eq. \eqref{eq_TIquad}, here the only difference is that the expectation at $\beta=0$ is approximated by using samples from the restricted posteriors, $\pi_+(\x)$ and $\pi_-(\x)$, instead of the posterior $\pi(\x)$.\footnote{In order to obtain samples from  $\pi_\pm(\x)$, we just need to consider   $\pi_\pm(\x)$ as target density instead of $\pi(\x)$, in the MCMC steps. A similar alternative procedure is to apply rejection sampling, discarding the samples from $\pi(\x)$ such that $f_\pm(\x)=0$.}
To obtain an approximation of the true quantities of interest $\eta_+$, $\eta_-$ (instead of $\eta^\text{res}_+$ and $\eta^\text{res}_-$), we compute two correction factors from a single set of $K$ samples from $\post(\x)$ as follows
\begin{align}
	\widehat{R}_+= \frac{1}{K}\sum_{i=1}^K\mathbbm{1}_{\mathcal{X}_+}(\z_i)&\approx \frac{Z_+}{Z},   \\
	\widehat{R}_- = \frac{1}{K}\sum_{i=1}^K\mathbbm{1}_{\mathcal{X}_-}(\z_i)&\approx \frac{Z_-}{Z}, \quad \z_i \sim \post(\x),
\end{align}
where $\mathbbm{1}_{\mathcal{X}_+}(\x_i)=1$ if $f_+(\x_i)>0$, $\mathbbm{1}_{\mathcal{X}_-}(\x_i)=1$ if $f_-(\x_i)>0$, and both zero otherwise. 
The final estimator of $I$ is
\begin{align}\label{eq_estimCorrect}
	\widehat{I} = \widehat{R}_+\exp\left(\widehat{\eta}^\text{res}_+\right) -\widehat{R}_-\exp\left(\widehat{\eta}^\text{res}_-\right),
\end{align}
including the two correction factors. Table \ref{table_generic_f} provides all the details of GTI in this scenario.

\begin{table}[!ht]
	\caption{\textbf{GTI for generic functions $f(\x)$}} 	\label{table_generic_f}
	\vspace{-0.4cm}
	\begin{tabular}{|p{0.95\columnwidth}|}
		\hline
		\vspace{0.1cm}
		{\bf - Initialization:} Choose the set of nodes $\{\beta_i\}_{i=1}^N$ (with $\beta_1=0$ and $\beta_N=1$), and the number of iterations $N$.
		\newline
		{\bf - For $i=1,\ldots,N$:}
		\begin{enumerate}
			\item  {\bf Sampling:} Sample 
			\begin{align*}
				\{\x_{i,m}\}_{m=1}^M\sim\bar{\varphi}_+(\x|\beta_i)\propto f_+(\x)^{\beta_i}\pi_+(\x)
				,\quad \pi_+(\x)=\pi(\x)
				\mathbbm{1}_{\mathcal{X_+}}(\x), \\
				\{{\bf v}_{i,m}\}_{m=1}^M\sim \bar{\varphi}_-(\x|\beta_i)\propto f_-(\x)^{\beta_i}\pi_-(\x),\quad \pi_-(\x)=\pi(\x)\mathbbm{1}_{\mathcal{X_-}}(\x).
			\end{align*}
			\item {\bf Compute:} 
			\begin{align*}
			\widehat{E}^+_i = \frac{1}{M}\sum_{m=1}^{M}\log f_+(\x_{i,m}), \qquad \widehat{E}^-_i = \frac{1}{M}\sum_{m=1}^{M}\log f_-({\bf v}_{i,m}).
			\end{align*}
			\item {\bf Update:} If $i=1$, set $\eta_{+}^{(i)}=\eta_{-}^{(i)}=0$ . If $i>1$, update recursive estimates
			\begin{align*}
			\widehat{\eta}_{+}^{(i)} \leftarrow \widehat{\eta}_{+}^{(i-1)} + 0.5\cdot(\beta_{i}-\beta_{i-1})\left(\widehat{E}^+_i+\widehat{E}^+_{i-1}\right), \\
			\widehat{\eta}_{-}^{(i)} \leftarrow \widehat{\eta}_{-}^{(i-1)} + 0.5\cdot(\beta_{i}-\beta_{i-1})\left(\widehat{E}^-_i+\widehat{E}^-_{i-1}\right).
			\end{align*}
			
		\end{enumerate}
		{\bf - Correction:} Compute correction factor using samples $\{\z_k\}_{k=1}^K\sim \post(\x)$,
		\begin{align*}
	\widehat{R}_+= \frac{1}{K}\sum_{i=1}^K\mathbbm{1}_{\mathcal{X}_+}(\z_i), \quad 
	\widehat{R}_- = \frac{1}{K}\sum_{i=1}^K\mathbbm{1}_{\mathcal{X}_-}(\z_i), \quad \z_i \sim \post(\x),
\end{align*}
		\newline
		{\bf - Outputs:} The final estimator  
	\begin{align*}
	\widehat{I} = \widehat{R}_+\exp\left(\widehat{\eta}^{(N)}_+\right) -\widehat{R}_-\exp\left(\widehat{\eta}^{(N)}_-\right),
	\vspace{-0.5cm}
\end{align*} \\ 
		\hline 	
	\end{tabular}

\end{table}

{\rem\label{Rem2}  {\bf Standard TI as special case of GTI:} Note that the GTI scheme contains TI as a special case if we set $f(\x)=\ell(\y|\x)$ (i.e., the likelihood function) and let the prior $g(\x)$ play the role of $\pi(\x)$. Since the likelihood $\ell(\y|\x)$ is non-negative we have $\eta_-=-\infty$ (then, $\exp\left(\eta_-\right)=0$), hence we only have to consider the estimation of $\eta_+$.
Moreover, if $\ell(\y|\x)$ is strictly positive we do not need to compute the correction factor.}
{\rem
The GTI procedure, described above, also allows the application of the standard TI for computing marginal likelihoods when the likelihood function is not strictly positive, by applying  a correction factor in the same fashion (in this case, considering a restricted prior pdf).  }

\section{Computational considerations and other extensions}\label{sec_compDet}

In this section, we discuss computational details, different scenarios and further extensions, that are listed below. 
\newline
{\bf Acceleration schemes.}  In order to apply GTI, the user must set $N$ and $M$, so that the total number of samples/evaluations of $f(\x)$ in Table \ref{table_f_stricly_pos} is $E=NM$. The evaluations of $f(\x)$ in Table \ref{table_generic_f} are $E=2NM + K$. We can reduce the cost of algorithm in Table \ref{table_generic_f} to $E=NM + K$ with an acceleration scheme.
Instead of running separate MCMC algorithms for $\bar{\varphi}_+(\x|\beta) \propto f_+(\x)^\beta \pi_+(\x)$ and $\bar{\varphi}_-(\x|\beta) \propto f_-(\x)^\beta \pi_-(\x)$, we use a single run targeting 
\begin{equation}
\bar{\varphi}_\text{abs}(\x|\beta) \propto |f(\x)|^\beta \pi(\x)\mathbbm{1}\left(f(\x)\neq 0\right).
\end{equation}
We can obtain two MCMC samples, one from $\bar{\varphi}_+(\x|\beta)$ and one from $\bar{\varphi}_-(\x|\beta)$, by separating the sample into two: samples with positive value of $f(\x)$, and samples with negative value of $f(\x)$, respectively. The procedure can be repeated until obtaining the desired number of samples from each density, $\bar{\varphi}_+(\x|\beta)$ and $\bar{\varphi}_-(\x|\beta)$.
\newline
Moreover, note that in Table \ref{table_generic_f} we need to draw samples from $\pi_+(\x)$, $\pi_-(\x)$ and $\pi(\x)$. 
Instead of sampling each one separately, we can use the following procedure.
Obtain a set of samples from $\pi(\x)$ and then apply rejection sampling (i.e. discard samples with $f_\pm(\x)=0$) in order to obtain samples from $\pi_\pm(\x)$.
Combining this idea with the acceleration scheme above reduces the cost of Table \ref{table_generic_f} to $E=MN$.
\newline
{\bf Parallelization.} 
Note that steps 1 and 2 in Table \ref{table_f_stricly_pos} and Table \ref{table_generic_f} are amenable to parallelization. In other words, those steps need not be performed sequentially but can be done using embarrassingly parallel MCMC chains (i.e. with no communication among $N$, or $2N$, workers). Only step 3 requires communicating to a central node and combining the estimates. 
With this procedure, the number of evaluations $E$ is the same but the computation time is reduced by $\frac{1}{N}$ (or $\frac{1}{2N}$) factor.
On the other hand, population MCMC techniques can be used, but parallelization speedups are lower since communication among workers occurs every so often, in order to foster the exploration of the chains \cite{martino2016orthogonal,calderhead2009estimating}.
\newline
{\bf Vector-valued functions ${\bf f}(\x)$.} In Bayesian inference, one is often interested in computing moments of the posterior, i.e.,
\begin{equation}
{\bf I}=\int_{\mathcal{X}} \x^{\alpha} \post(\x) d\x, \qquad \alpha \geq 1.
\end{equation}
 In this case ${\bf I}$ is a vector and ${\bf f}(\x)=\x^{\alpha}$. When $\alpha=1$,   ${\bf I}$ represents the {\it minimum mean square error} (MMSE) estimator. More generally, we can have a vector-valued function, 
$$
{\bf f}(\x) = \left[f_1(\x),\dots,f_{d_f}(\x)\right]^\top: \mathcal{X}\to \mathbb{R}^{d_f},
$$
hence the integral of interest is a vector ${\bf I}=[I_1,\dots,I_{d_f}]^\top$ where $I_i = \int_\mathcal{X}f_i(\x)\post(\x)d\x$.
In this scenario, we need to apply the GTI scheme to each component of ${\bf I}$ separately, obtaining estimates $\widehat{I}_i$ of the form in Eq. \eqref{eq_estimCorrect}. 
\newline
{\bf TI within the TABI framework: TATI.} 
We have seen that we can apply GTI to compute the posterior expectation of a generic $f(\x)$, that can be positive, negative and have zero-valued regions. For doing this, we connected  with a tempered path, $\pi_+(\x)$ and $\pi_-(\x)$, to $f_+(\x)\pi(\x)$ and $f_-(\x)\pi(\x)$ respectively and then apply correction factors.
\newline
An alternative procedure is to use the TABI identity in Eq. \eqref{eq:TABIidentity}, rather than Eq. \eqref{eq_3_const} and use {\it reference} distributions for computing separately $c_+$, $c_-$ and $Z$ in Eq. \eqref{eq_3_const}.
This target-aware TI (TATI) differ from GTI in that we need to apply TI three times, and bridge three reference distributions to the target densities $f_+(\x)\pi(\x)$, $f_-(\x)\pi(\x)$ and $\pi(\x)$.
Let us define as
$$
p^\text{ref}_i(\x), \qquad i=1,2,3,
$$
three unnormalized {\it reference} densities with normalizing constants,
$$
Z^\text{ref}_i=\int_{\mathcal{X}} p^\text{ref}_i(\x) d\x, \qquad i=1,2,3.
$$
Then, the idea is to apply TI for obtaining estimates of $\log\frac{c_+}{Z^\text{ref}_1}$, $\log\frac{c_-}{Z^\text{ref}_2}$ and $\log\frac{Z}{Z^\text{ref}_3}$.
A requirement is that $p^\text{ref}_1(\x)$ is zero where $f_+(\x)\pi(\x)$ is zero,  $p^\text{ref}_2(\x)$  is zero where $f_-(\x)\pi(\x)$ is zero, and $p^\text{ref}_3(\x)$ is zero where $\pi(\x)$ is zero. Namely, we need to be able to build a continuous path between the reference distributions and the corresponding unnormalized pdf of interest. With this procedure, we do not need to apply correction factors, but we just need to apply the algorithm in Table \ref{table_f_stricly_pos} three times.
The performance of TATI is expected to be better than GTI if we are able to choose three reference distributions that are `closer' to the corresponding target densities, than what $\pi(\x)$ is to $f_+(\x)\pi(\x)$ or $f_-(\x)\pi(\x)$ \cite{llorente2020marginal}. For instance, we can obtain the reference pdfs by building nonparametric approximations to each target density \cite{llorente2021deep}. 

\section{Numerical experiments}\label{sec_exper}

In this section, we illustrate the performance of the proposed scheme in two numerical experiments which consider different kind of densities $\post$ with different features and different dimensions, and also different function $f(\x)$. 
In the first example, $f(\x)$ is strictly positive so we apply the algorithm described in Table \ref{table_f_stricly_pos}. In the second example, we consider $f(\x)$ to have zero-valued regions, and hence we apply the algorithm in Table \ref{table_generic_f}.
Notice we consider the same setup as in \cite{rainforth2020target} in order to compare with respect to instances of TABI algorithms.

\subsection{First numerical analysis}\label{sec_first_exp}
Let us consider the following Gaussian model \cite{rainforth2020target}
\begin{align}
g(\x) = \mathcal{N}(\x|\mathbf{0}_D, \mathbf{I}_D),\quad \ell(\y|\x) = \mathcal{N}\left(-\frac{y}{\sqrt{D}}\mathbf{1}_D \Big| \x, \mathbf{I}_D\right), \quad f(\x) = \mathcal{N}\left(\x \Big|  \frac{y}{\sqrt{D}}\mathbf{1}_D, \frac{1}{2}\mathbf{I}_D\right),
\end{align}
where $D$ is the dimensionality, $\mathbf{I}_D$ is the identity matrix, $\mathbf{0}_D$ and $\mathbf{1}_D$ are $D$-vectors containing only zeros or ones respectively, and $y$ is a scalar value that represents the radial distance of the observation $\y=-\frac{y}{\sqrt{D}}\mathbf{1}_D$ to the origin. 
We are interested in the estimation of $I=\int_{\mathcal{X}} f(\x)\post(\x)d\x$.
Thus this problem consists in computing the posterior predictive density, under the above model, at the point $\frac{y}{\sqrt{D}}\mathbf{1}_D$.
In this toy example, the posterior and the function-scaled posteriors can be obtained in closed-form, that is,
\begin{align}\label{eq_tempered_Ex1}
	\bar{\varphi}(\x|\beta)= \mathcal{N}\left(\x \Big|  \frac{2\beta - 1}{2\beta + 2}\frac{y}{\sqrt{D}}\mathbf{1}_D, \frac{1}{2\beta+2}\mathbf{I}_D\right),\quad \beta\in[0,1],
\end{align}
and 
\begin{align}
	\post(\x) = \mathcal{N}\left(\x \Big|  -\frac{1}{2}\frac{y}{\sqrt{D}}\mathbf{1}_D, \frac{1}{2}\mathbf{I}_D\right).
\end{align}
The ground-truth is known, and can be written as a Gaussian density evaluated at $\frac{y}{\sqrt{D}}\mathbf{1}_D$, more specifically,
$I = \mathcal{N}\left(\frac{y}{\sqrt{D}}\mathbf{1}_D \Big| -\frac{1}{2}\frac{y}{\sqrt{D}}\mathbf{1}_D, \mathbf{I}_D \right)$.
\newline
We test the values $y \in \{2,3.5,5\}$ and $D \in \{10,25,50\}$. Note that, as we increase $y$, the posterior $\post(\x)$ and the density $\bar{\varphi}(\x|1)\propto f(\x)\pi(\x)$ become further apart.

\begin{figure}[h!]
	\centering
	\begin{subfigure}[b]{0.31\textwidth}
		\includegraphics[width=1\textwidth]{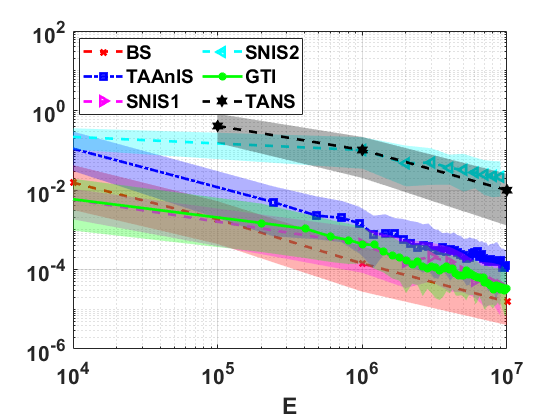}
				\caption{$y=2$, $D=10$}
	\end{subfigure}
	\begin{subfigure}[b]{0.31\textwidth}
		\includegraphics[width=1\textwidth]{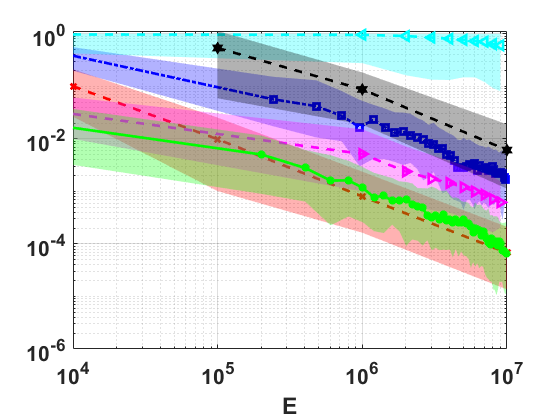}
		\caption{$y=2$, $D=25$}
	\end{subfigure}
	\begin{subfigure}[b]{0.31\textwidth}
		\includegraphics[width=1\textwidth]{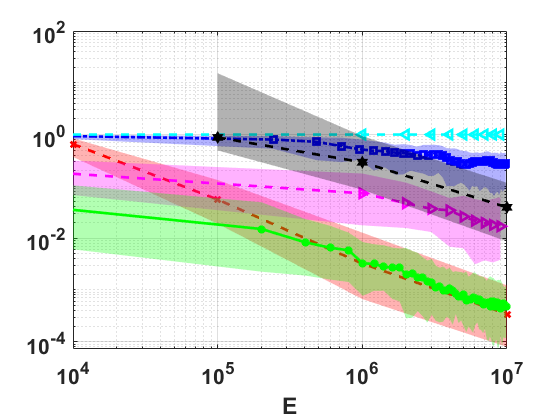}
		\caption{$y=2$, $D=50$}
	\end{subfigure}
	\begin{subfigure}[b]{0.31\textwidth}
		\includegraphics[width=1\textwidth]{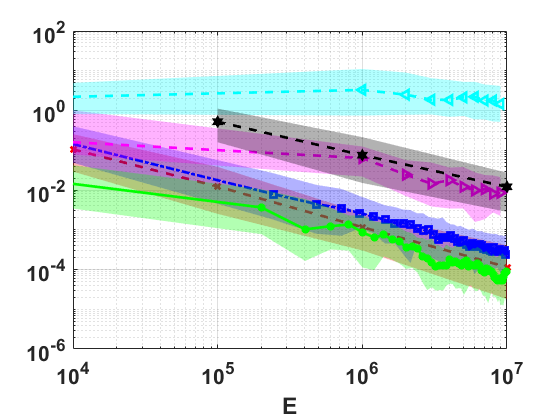}
				\caption{$y=3.5$, $D=10$}
	\end{subfigure}
	\begin{subfigure}[b]{0.31\textwidth}
		\includegraphics[width=1\textwidth]{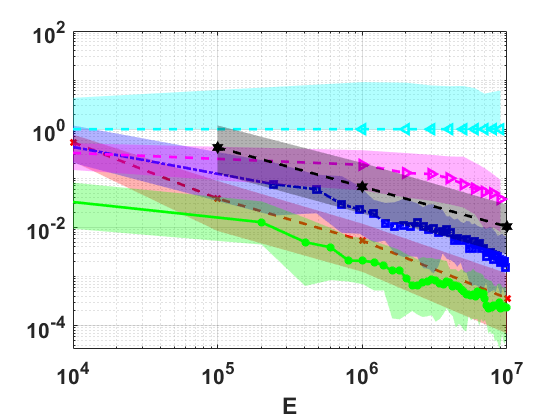}
		\caption{$y=3.5$, $D=25$}
	\end{subfigure}
	\begin{subfigure}[b]{0.31\textwidth}
		\includegraphics[width=1\textwidth]{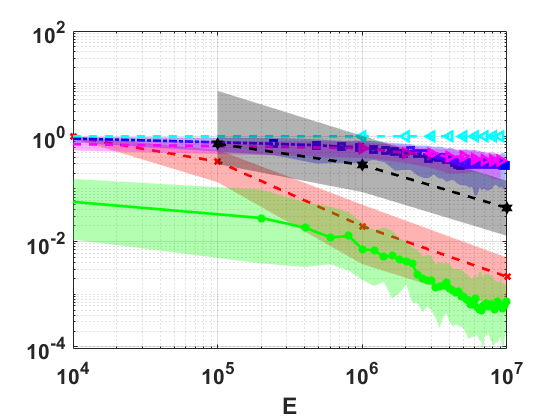}
		\caption{$y=3.5$, $D=50$}
	\end{subfigure}
	\begin{subfigure}[b]{0.31\textwidth}
		\includegraphics[width=1\textwidth]{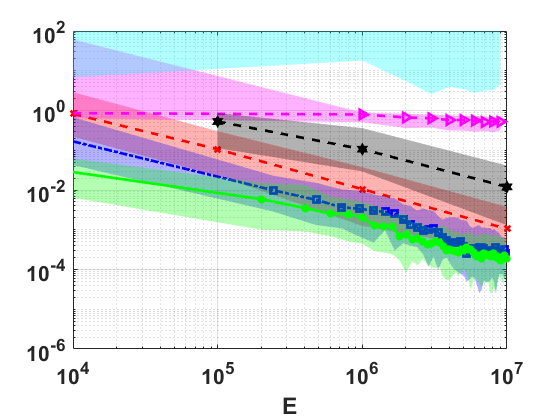}
				\caption{$y=5$, $D=10$}
	\end{subfigure}
	\begin{subfigure}[b]{0.31\textwidth}
		\includegraphics[width=1\textwidth]{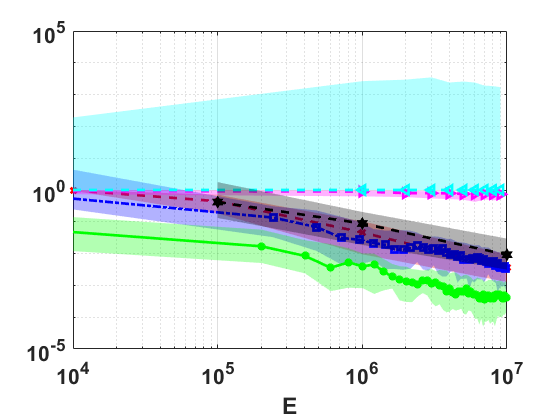}
				\caption{$y=5$, $D=25$}
	\end{subfigure}
	\begin{subfigure}[b]{0.31\textwidth}
	\includegraphics[width=1\textwidth]{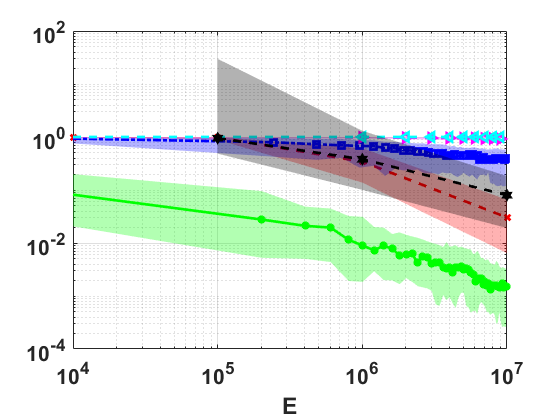}
			\caption{$y=5$, $D=50$}
	\end{subfigure}
	\caption{\label{fig_gaussianEx} Relative squared error of the considered algorithms as a function of number of total likelihood evaluations $E$, for different $y$ and $D$.	The median, 25\% and 75\% quantiles (over 100 independent simulations) are depicted.	
	}
\end{figure}

{

\subsubsection{Comparison with other target-aware approaches}

We aim to compare GTI with a MCMC baseline and other target-aware algorithms, that make use of $f(\x)$. 
Specifically, we compare against two extreme cases of self-normalized IS (SNIS) estimators $\widehat{I}_{\text{SNIS}} =\frac{1}{\sum_{j=1}^{M_\text{tot}}w_j} \sum_{i=1}^{M_\text{tot}}w_i f(\x_i)$, where $\x_i\sim q(\x)$ and $w_i = \frac{\pi(\x_i)}{q(\x_i)}$ is the IS weight.  
Namely, {\bf (1)} SNIS using samples from the posterior (SNIS1), i.e.,  $q(\x) = \post(\x)$ (hence, SNIS1 coincides with MCMC),
and {\bf (2)} SNIS using samples from  $q(\x) = \bar{\varphi}(\x|1)\propto f(\x)\pi(\x)$ (SNIS2), which corresponds to setting $\beta=1$ in Eq. \eqref{eq_tempered_Ex1}.
These choices are optimal for estimating, respectively, the denominator and the numerator of the right hand side of Eq. \eqref{eq:IntegralOfInter} \cite{Robert04}.
Note that SNIS2 can be considered as a first ``primitive'' target-aware algorithm, since it employs samples from $\bar{\varphi}(\x|1) \propto f(\x)\pi(\x)$.
\newline
A second target-aware approach can be obtained by recycling the samples generated in SNIS1 and SNIS2 (that is, from $\post(\x)$ and $\bar{\varphi}(\x|1)$), and is called {\bf (3)} bridge sampling (BS) \cite{llorente2020marginal}.  This estimator can be viewed as if we use the mixture of $\post(\x)$ and $\bar{\varphi}(\x|1)$ as proposal pdf.
More details about (optimal) BS can be found in App. \ref{App:BridgeSampling}.
Finally, we also aim to compare against the target-aware versions of two popular marginal likelihood estimators, namely, {\bf (4)} target-aware annealed IS (TAAnIS) and {\bf (5)} target-aware nested sampling (TANS), that also make use of the $f(\x)$ \cite{rainforth2020target}. 
\newline
In order to keep the comparisons fair, we consider the same number of likelihood evaluations $E$ in all of the methods. Note that evaluating the likelihood is usually the most costly step in many real-world scenarios. 
Hence, in SNIS1 and SNIS2 we draw $M_\text{tot} = E$ samples from $\post(\x)$ and $\bar{\varphi}(\x|1)$, respectively, via MCMC; BS employ $\frac{E}{2}$ samples from $\post(\x)$ and $\frac{E}{2}$ from $\bar{\varphi}(\x|1)$. 
For TAAnIS and TANS we use the same parameters as in \cite{rainforth2020target}. Namely, for TAAnIS we employ $N=200$ intermediate distributions, with $n_\text{MCMC}= 5$ iterations of Metropolis-Hastings (MH) algorithm, which allow for a total number of particles $n_\text{par} = \lfloor \frac{E}{(N-1)n_\text{MCMC} + N-1} \rfloor$, where half of the particles are used to estimate the numerator, and the other half to estimate the denominator on the right-hand side of Eq. \eqref{eq:IntegralOfInter}. For TANS, we employ $n_\text{MCMC}= 20$ iterations of MH and $n_\text{par} = \lfloor \frac{E}{1 + \lambda n_\text{MCMC}} \rfloor$ particles, where $\lambda = 250$ and $T = \lambda n_\text{par}$ iterations. Again, TANS employs one half of the particles for estimating the numerator and the other half for the denominator. 
Finally, in GTI we set also $N=200$, hence we draw $M = \lfloor \frac{E}{N}\rfloor$ samples from each $\bar{\varphi}(\x|\beta_i),\ i=1,\dots,N$. Note that we are setting the same number of intermediate distributions in GTI and TAAnIS, however, the paths are not identical, since TAAnIS aims at bridging the prior with $\pi(\x)$ and $\varphi(\x|1)$, while GTI directly bridges $\pi(\x)$ with $\varphi(\x|1)$. All the iterations of the MH algorithm use a Gaussian random-walk proposal with covariance matrix equal to $\Sigma = 0.1225 {\bf I}$, $\Sigma = 0.04 {\bf I}$ and $\Sigma = 0.01 {\bf I}$, for $D=10,25,50$ respectively. Following \cite{rainforth2020target}, for TANS we use instead  $\Sigma = {\bf I}$, $\Sigma = 0.09 {\bf I}$ and $\Sigma = 0.01 {\bf I}$. 
For choosing $\beta_i$ in TAAnIS and GTI, we use the powered  fraction schedule, $\beta_i = \left(\frac{i-1}{N-1}\right)^5$ for $i=1,\dots,N$ \cite{friel2008marginal,xie2010improving}.

%
%
%

}

{

\subsubsection{Results}

The results are given in Figure \ref{fig_gaussianEx}, which show, for each pair $(y,D)$, the median relative square error along with the 25\% and 75\% quantiles (over 100 simulations) versus the number of total likelihood evaluations $E$, up to $E=10^7$.
We see that GTI is the first or second best overall, in terms of relative squared error for all $(y,D)$. 
In fact, the performance of GTI seems rather insensitive to increasing the dimension $D$ and $y$.
We see that for low dimension and when the distance  between $\post(\x)$ and $\bar{\varphi}(\x|1)$ is small (i.e. $y=2$ and $D=10$), the target-aware algorithms do not produce large gains with respect to the MCMC baseline (SNIS1). 
On the contrary, for $y=3.5,5$ (second and third row), we see that the target-aware algorithms, GTI, TAAnIS and BS, outperform the MCMC baseline. 
This performance gain with larger $y$ is expected, since this represents a larger mismatch between $\post(\x)$ and $\bar{\varphi}(\x|1)\propto f(\x)\pi(\x)$, which is a scenario where the target-aware approaches are well suited. 
Comparing the target-aware algorithms, we see that TAAnIS performs also as well as our GTI in low dimensions ($D=10$), but it breaks down as we increase the dimension, being outperformed by TANS in $D=25,50$, confirming the results of \cite{rainforth2020target} where TANS is preferable over TAAnIS in high dimension.
It is worth noticing the very good performance of BS, given its simplicity and that it can be computed at almost no extra cost once we have computed SNIS1 and SNIS2.
Indeed, its performance matches that of GTI, and actually outperforms GTI when the separation is not too high. 
This is also expected since, when $y=2$, both $\post(\x)$ and $\bar{\varphi}(\x|1)$ are good pdfs for estimating both numerator and denominator of the right hand side of Eq. \eqref{eq:IntegralOfInter}. 
In this sense, having only one ``bridge'' is better than having $n=200$ intermediate distributions.
However, GTI outperforms BS when $y=3.5,5$, especially when the dimension is high. 
\newline
In summary,
our proposed GTI is able to produce good estimates in the range of values of $(y,D)$ considered. The performance gains with respect to a MCMC baseline are higher when the discrepancy between $\post(\x)$ and $\bar{\varphi}(\x|1)$ is large. As compared to other target-aware approaches, GTI produce better estimates (especially in high dimension) and is also able to perform well when the discrepancy is low, matching the performance of BS, that is a simpler and more direct target-aware algorithm. 



}

\subsection{Second numerical analysis}

We consider the following two-dimensional banana-shaped density (which is a benchmark example \cite{rainforth2020target,LucaJesse2,haario2001adaptive}),
\begin{align}
	\pi(x_1,x_2) = \exp\left(-\frac{1}{2}\left(0.03x_1^2 + \left(\frac{x_2}{2}+0.03\left(x_1^2-100\right)\right)^2\right)\right) \cdot \mathbbm{1}\left( \x\in \mathcal{B}\right),
\end{align}
where $\mathbbm{1}\left( \x\in \mathcal{B}\right)$ is the prior, where $\mathcal{B} = \{\x:\ -25 < x_1 < 25,\ -40 < x_2 <20 \}$,
and the function is
\begin{align}
	f(x_1,x_2) = (x_2+10)\exp\left(-\frac{1}{4}\left(x_1+x_2+25\right)^2\right)\mathbbm{1}\left(x_2>-10\right).
\end{align}
We compare GTI using $N\in\{10,50, 100\}$ against TAAnIS and TANS, in the estimation of $\E_{\post}[f(\x)]$ allowing a budget of $E=10^6$. We also consider a baseline MCMC chain targeting $\pi(\x)$ with the same number of likelihood evaluations. 
\newline
{
The main difference with respect to previous experiment is that $f(\x)$ here has a zero-valued region, so, in order to apply GTI, we need to use the algorithm in Table \ref{table_generic_f}. Hence, for GTI, we run $N+1$ chains for $M=\frac{E}{N+1}$ iterations. The first $N$ chains address a different tempered distribution $f(\x)^{\beta_i}\pi(\x)$, and the last chain is used to compute the correction factor. 
It is also important to notice here that TAAnIS, which also uses a geometric path to bridge $f(\x)\pi(\x)$ and the prior, requires also the computation of a correction factor, accounting for the fact we connect a prior restricted to where $f(\x)\neq 0$. This amounts to multiply by a factor $\frac{1}{2}$ the final estimate returned by TAAnIS.
\newline
All the MCMC algorithms use a Gaussian random-walk proposal with covariance $\Sigma=3\mathbf{I}_2$. The budget of likelihood evaluations is $E=10^6$, for all the compared schemes. We use again the powered  fraction schedule: $\beta_i = \left(\frac{i-1}{N-1}\right)^5$ for $i=1,\dots,N$.
\newline
{\bf Results.} The results are shown in Table \ref{table_1}. We show the median relative square error of the methods over 100 independent simulations. 
For the sample size considered, GTI performs better than MCMC baseline and the other target-aware algorithms. 
TAAnIS performs slightly better than the MCMC baseline, while TANS completely fails at estimating the posterior expectation in this example.
For $N=100$, the performance gains of GTI are almost of one order of magnitude over MCMC. 
However, note that GTI with the choice $N=10$ is worse than the MCMC baseline due to the discretization error, i.e., there are not enough quadrature nodes, so the estimation in Eq. \eqref{eq_TIquad} has considerable bias. In that situation, increasing the sample size would not translate into a significant performance gain.
This contrasts with TAAnIS, where increasing $N$ produces only small improvement on the final estimate, since TAAnIS is unbiased regardless of the choice of $N$. 
}

\begin{figure}[h!]
	\centering
	\begin{subfigure}[b]{0.31\textwidth}
		\includegraphics[width=1\textwidth]{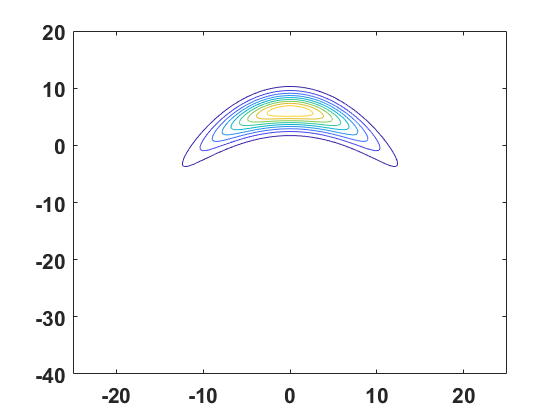}
		\caption{$\pi(\x)$}
	\end{subfigure}
	\begin{subfigure}[b]{0.31\textwidth}
		\includegraphics[width=1\textwidth]{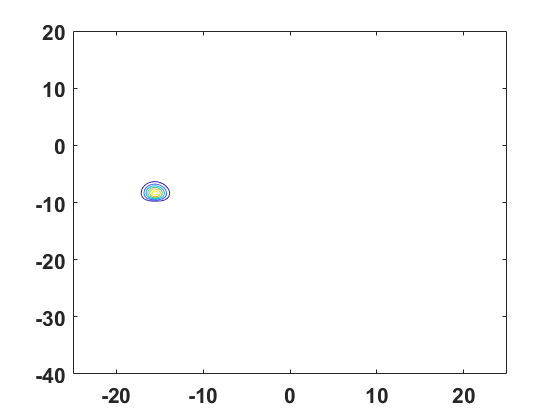}
		\caption{$f(\x)\cdot\pi(\x)$}
	\end{subfigure}
	\begin{subfigure}[b]{0.31\textwidth}
		\includegraphics[width=1\textwidth]{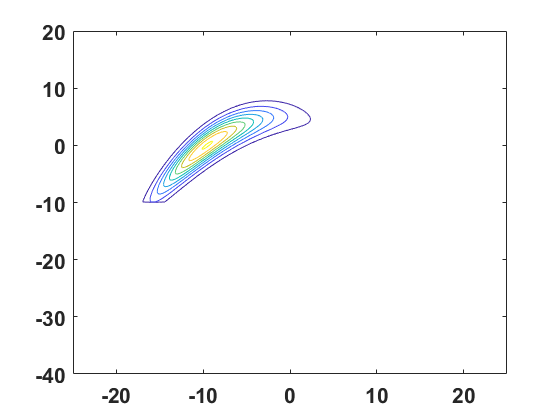}
		\caption{$f(\x)^{\beta}\cdot\pi(\x)$}
	\end{subfigure}
	\caption{\label{fig_banana} Plots of $\pi(\x)$, $f(\x)\pi(\x)$ and $f(\x)^\beta \pi(\x)$ with $\beta=0.0173$ for the banana example. We see that $f(\x)$ and $f(\x)\pi(\x)$ have little overlap and hence a direct MCMC estimate of $\E_{\post}[f(\x)]$ is not efficient. 
		The tempered distributions, $f(\x)^\beta \pi(\x)$, are in-between those distributions, helping in the estimation of $\E_{\post}[f(\x)]$.
	}
\end{figure}

\begin{table*}[!ht]
	\centering
	\caption{
		Median relative square error of GTI, TAAnIS and TANS for $E=10^6$ likelihood evaluations and $N\in\{10,50,100\}$. Best result is boldfaced.	
	}

	\begin{tabular}{|c|c|c|}
	\hline	
	Algorithm & $N$ & Relative SE \\
	\hline
	\hline
	MCMC & $--$ & 0.0040054 \\
	\hline
	GTI  & 10 & 0.01516  \\
	GTI & 50 & 0.0012224 \\
	GTI & 100 & $\mathbf{0.00060778}$ \\
	\hline
	TAAnIS & 10  & 0.0037181 \\
	TAAnIS & 50  & 0.0036433  \\
	TAAnIS & 100  & 0.0033898 \\
	\hline
	TANS & $--$ &  0.07276 \\
	\hline
\end{tabular}

	\label{table_1}
\end{table*}


%
%
%
%
%
%
%

\section{Conclusions}\label{sec_conc}
We have extended the powerful thermodynamic integration technique for performing a target-aware Bayesian inference. Namely, GTI allows the computation of posterior expectations of real-valued functions $f(\x)$, and also vector-valued functions ${\bf f}(\x)$. GTI contains the standard TI as special case. Even for the estimation of the marginal likelihood, this work provides a way for extending the application of the standard TI avoiding the assumption of strictly positive likelihood functions (see Remarks \ref{Rem1}- \ref{Rem2}). 
Several computational considerations and variants are discussed. The advantages of GTI over other target-aware algorithms are shown in different numerical comparisons.
As a future research line, we plan to study new continuous paths for linking densities with different support, avoiding the need of the correction terms. 
Alternatively, as discussed in Sect. \ref{sec_compDet}, another approach would be to design suitable approximations of $\varphi_+(\x)$, $\varphi_-(\x)$ and $\pi(\x)$ (see the end of Sect. \ref{sec_compDet}) using, e.g., regression techniques \cite{llorente2021deep,llorente2020adaptive}.


\section*{Acknowledgments}

The work was partially supported by the Young Researchers R\&D Project, ref. num. F861 (AUTO-BA- GRAPH) funded by Community of Madrid and Rey Juan Carlos University, by Agencia Estatal de Investigaci\'on AEI (project SP-GRAPH, ref. num. PID2019-105032GB-I00), and by Spanish government via grant FPU19/00815.


\bibliographystyle{IEEEtranN}
\bibliography{bibliografia}

{
\begin{appendices}

\section{Bridge sampling}\label{App:BridgeSampling}

The estimator tested in Sect. \ref{sec_first_exp} is an instance of bridge sampling.
Bridge sampling (BS) is an importance sampling approach for computing the ratio of normalizing constants  of two unnormalized pdfs using samples from both densities \cite{llorente2021computation}. Here, the two unnormalized pdfs of interest are $\pi(\x)$ and $\varphi(\x)=f(\x)\pi(\x)$, and the ratio $\frac{c}{Z}$ corresponds to the posterior expectation of interest, namely, $I = \frac{c}{Z}$. 
Hence, BS can be viewed as a target-aware approach. 
In order to implement the optimal bridge sampling estimator, an iterative scheme is required.
\newline
Let 
$
\{\x_i\}_{i=1}^{N_1},\ 
\{\z_i\}_{i=1}^{N_2}
$
denote  sets of MCMC samples from $\post(\x)$ and $\bar{\varphi}(\x) = \frac{\varphi(\x)}{c}$.
Let $\widehat{I}^{(0)}$ be an initial estimate of $I$, the optimal BS estimator is computed by refining this estimate through the following loop. For $t=1,\dots,T$:
\begin{align}
\widehat{I}^{(t)} = \frac{\frac{1}{N_1}\sum_{i=1}^{N_1}\dfrac{\varphi(\x_i)}{N_2\varphi(\x_i) + N_1\widehat{I}^{(t-1)}\pi(\x_i)}}{\frac{1}{N_2}\sum_{i=1}^{N_2}\dfrac{\pi(\z_i)}{N_2\varphi(\z_i) + N_1\widehat{I}^{(t-1)}\pi(\z_i)}}.
\end{align} 
In the experiments,  just a couple of iterations were needed for $\widehat{I}^{(t)}$ to converge. 
As initial estimate, we take
\begin{align}
\widehat{I}^{(0)} &= \frac{1}{N_1}\sum_{i=1}^{N_1}f(\x_i).
\end{align}

\section{TI when $f(\x)$ has zero-valued regions}\label{App:ContinuousPath}

We have seen that the case of $f(\x)$ being strictly positive (or strictly negative) is apt for thermodynamic integration since $\E_{\bar{\varphi}(\x|\beta)}[\log f(\x)]<\infty$ for all $\beta$, so the integral in the r.h.s. of Eq. \eqref{eq_f_strictly} is finite. In other words, the integrand function, $\E_{\bar{\varphi}(\x|\beta)}[\log f(\x)]$, is continuous and bounded for $\beta\in[0,1]$. 
Now, we discuss the case where this does not hold. For instance, when $f(\x)$ has zero-valued regions within the support $\mathcal{X}$. In that case, the integrand of Eq. \eqref{eq_f_strictly} at $\beta=0$ diverges,
\begin{align}\label{eq_fer}
\E_{\bar{\varphi}(\x|0)}[\log f(\x)] = \int_\mathcal{X}\log f(\x)\post(\x)d\x = -\infty,
\end{align}
because the integral will sum over regions where $\log f(\x) = -\infty$. 
Note that the integrand
$$\E_{\bar{\varphi}(\x|0)}[\log f(\x)] = -\infty\ \text{at}\ \beta=0, \quad \text{but} \quad -\infty<\E_{\bar{\varphi}(\x|\beta)}[\log f(\x)] <\infty\ \text{for}\ \beta>0,$$
since the expectations,
\begin{align}
\E_{\bar{\varphi}(\x|\beta)}[\log f(\x)] \propto \int_\mathcal{X}\log f(\x)f(\x)^\beta\pi(\x)d\x,\quad \beta>0,
\end{align}
are w.r.t. densities that take into account $f(\x)$, and then the effective support is $\mathcal{X}\backslash \{\x:\ f(\x)=0\}$.
\newline
{\bf Improper integral.} In this case, the integral in Eq. \eqref{eq_f_strictly} is thus improper and has to be rewritten as
\begin{align}\label{eq_improper}
\eta = \lim_{\beta_0 \to 0^+}\int_{\beta_0}^1\E_{\bar{\varphi}(\x|\beta)}\left[\log f(\x)\right]d\beta.
\end{align}
If this limit exists, the integral is convergent and it is safe to apply quadrature (Riemann sums) to calculate it, taking a very small $\beta_0$.
\newline
{\bf Behavior near $\beta=0$.} 
Paying attention to the behavior of $\E_{\bar{\varphi}(\x|\beta)}[\log f(\x)]$ near $\beta=0$, we should notice that $\E_{\bar{\varphi}(\x|\beta)}[\log f(\x)]$ will not diverge to $-\infty$ as we get close to $\beta=0$. On the contrary, there is a lower limit on its value as we approach $\beta=0$. Consider, for an infinitesimal $\epsilon$, the integral
\begin{align}
\E_{p_\epsilon}[\log f(\x)] = \frac{\int_\mathcal{X}\log f(\x) f(\x)^\epsilon \pi(\x)d\x}{\int_\mathcal{X}f(\x')^\epsilon\pi(\x')d\x'}, \quad \epsilon \ll 1,
\end{align}
where $p_\epsilon(\x) \propto f(\x)^\epsilon\pi(\x)$ coincides with $\post(\x)$ in $\mathcal{X}\backslash \{\x: f(\x)=0\}$, and is different only in that $p_\epsilon(\x) =0$ whenever $f(\x)=0$, that is,
\begin{align}
p_\epsilon(\x)\approx \post_\text{res}(\x) = \frac{\post(\x)}{\int_{\mathcal{X}_0} \post(\x)d\x},\quad \x\in\mathcal{X}_0=\mathcal{X}\backslash\{\x:f(\x)=0\}.
\end{align}
This integral effectively corresponds to 
\begin{align}
\E_{p_\epsilon}[\log f(\x)] 
\approx \int_{\mathcal{X}_0}\log f(\x) \post_\text{res}(\x), \quad \epsilon \ll 1,
\end{align}
where $\mathcal{X}_0 = \mathcal{X}\backslash \{\x: f(\x)=0\}$, that is, the expectation is w.r.t. $\post_\text{res}(\x)$, the posterior restricted to regions where $f(\x)>0$. 
Hence, we can summarize this as follows:
\begin{align}
\text{As}\quad \beta\to0^+,\quad \E_{\bar{\varphi}(\x|\beta)}[\log f(\x)] \to \int_{\mathcal{X}_0}\log f(\x) \post_\text{res}(\x)d\x.
\end{align}
\newline
In summary, the integrand has a jump at $\beta=0$ since
\begin{align}
\E_{\bar{\varphi}(\x|\beta)}[\log f(\x)]\biggr\lvert_{\beta=0}=-\infty ,\quad \text{but} \quad  \lim_{\beta\to0^+}\E_{\bar{\varphi}(\x|\beta)}[\log f(\x)] = \int_{\mathcal{X}_0}\log f(\x) \post_\text{res}(\x)d\x.
\end{align}
Then, by using Eq. \eqref{eq_improper}, are we actually estimating
\begin{align}
\log \frac{\int f(\x)\pi(\x)d\x}{\int \pi_\text{res}(\x)d\x} = \log \frac{c}{Z_\text{res}},
\end{align}
instead of the integral of interest $I=\frac{c}{Z}$.
We need to apply a correction factor to our estimator as follows
\begin{align}
\widehat{\eta} + \log \frac{Z_\text{res}}{Z},
\end{align}
where the last term can be approximated from a posterior sample as follows:
\begin{align}
\frac{Z_\text{res}}{Z} \approx \frac{1}{N}\sum_{i=1}^N \mathbbm{1}_{\mathcal{X}_0}(\x_i),\quad \x_i\sim\post(\x),
\end{align}
where $\mathbbm{1}_{\mathcal{X}_0}(\x)$ is the indicator function in $\mathcal{X}_0$, i.e., $\mathbbm{1}_{\mathcal{X}_0}(\x_i)=1$ if $f(\x_i)>0$ and zero otherwise. 

\end{appendices}

}

\end{document}